\documentstyle[11pt,newpasp,twoside,epsf]{article}
\reversemarginpar

\begin{document}
\boldmath
\title{Constraints on $(\Omega_{\mathrm{m}},\Omega_\lambda)$
from strong lensing clusters}
\unboldmath
 \author{Ghislain Golse, Jean-Paul Kneib \& Genevi\`eve Soucail}
\affil{Laboratoire d'Astrophysique, 14, av. E.-Belin, F-31400 Toulouse, France}

\begin{abstract}
We use three strong lensing clusters to constrain the cosmological parameters
$\Omega_{\mathrm{m}}$ and $\Omega_\lambda$. Recent HST observations of galaxy
clusters reveal a large number of multiple images, which are predicted to be 
at different redshifts. We showed in a previous work that if it is possible to
measure spectroscopically the redshift of many multiple images then one can 
constrain $(\Omega_{\mathrm{m}},\Omega_\lambda)$ through ratios of
angular diameter distances independently of any external assumptions. 
Using three
strong lensing clusters, our combined results lead to tight constraints.
\end{abstract}

\section{Introduction}

Recent work on constraining the cosmological parameters using the CMB 
and high redshift supernovae seems to 
converge to a ``standard cosmological 
model'' favouring a flat universe with $\Omega_{\mathrm{m}}\simeq0.3$ and 
$\Omega_\lambda\simeq0.7$ (Jaffe et al. 2001). However, these 
results are still uncertain and it is 
therefore important to explore other independent techniques to constrain these
cosmological parameters.

The existence of multiple images -- with known redshifts -- of the same 
source allows to calibrate in an absolute way the total cluster mass deduced 
from the lens model. The great improvement in the mass modelling of 
cluster-lenses leads to the conclusion that clusters can also be used to 
constrain the geometry of the universe, through the ratio of angular size 
distances, $E=D_{\mathrm{LS}}/D_{\mathrm{OS}}$ (O: observer, L: lens, S: 
source), which only depends on the redshifts of the lens and sources, as well 
as the cosmological parameters (Link \& Pierce 1998).

\section{Method and Application}

In a previous work (Golse, Kneib, \& Soucail 2001b), we minimized a $\chi^2$ 
in the source planes to recover some parameters of the lens potential on a 
grid $(\Omega_{\mathrm{m}},\Omega_\lambda)$, using numerical simulations.
We apply this method to 3 strong lensing clusters which show several systems 
of multiple images with determined redshifts:

$\bullet$ AC114: $z_{\mathrm{L}}=0.312$, $z_{\mathrm{1}}=1.691$ 
(4 images), $z_{\mathrm{2}}=1.867$ (3 images), $z_{\mathrm{3}}=3.347$ 
(5 images) (Campusano et al. 2001),

$\bullet$ A2218: $z_{\mathrm{L}}=0.175$, $z_{\mathrm{1}}=0.702$ 
(4 images), $z_{\mathrm{2}}=1.034$ (3 images), $z_{\mathrm{3}}=2.515$ 
(3 images) (Ebbels et al. 1998),

$\bullet$ A1689: $z_{\mathrm{L}}=0.184$, $z_{\mathrm{1}}=1.834$ 
(4 images), $z_{\mathrm{2}}=4.868$ (2 images) (Golse et al. 2001a).

Fig.~1 shows the confidence levels on the cosmological parameters.

\begin{figure}[ht]
\plotfiddle{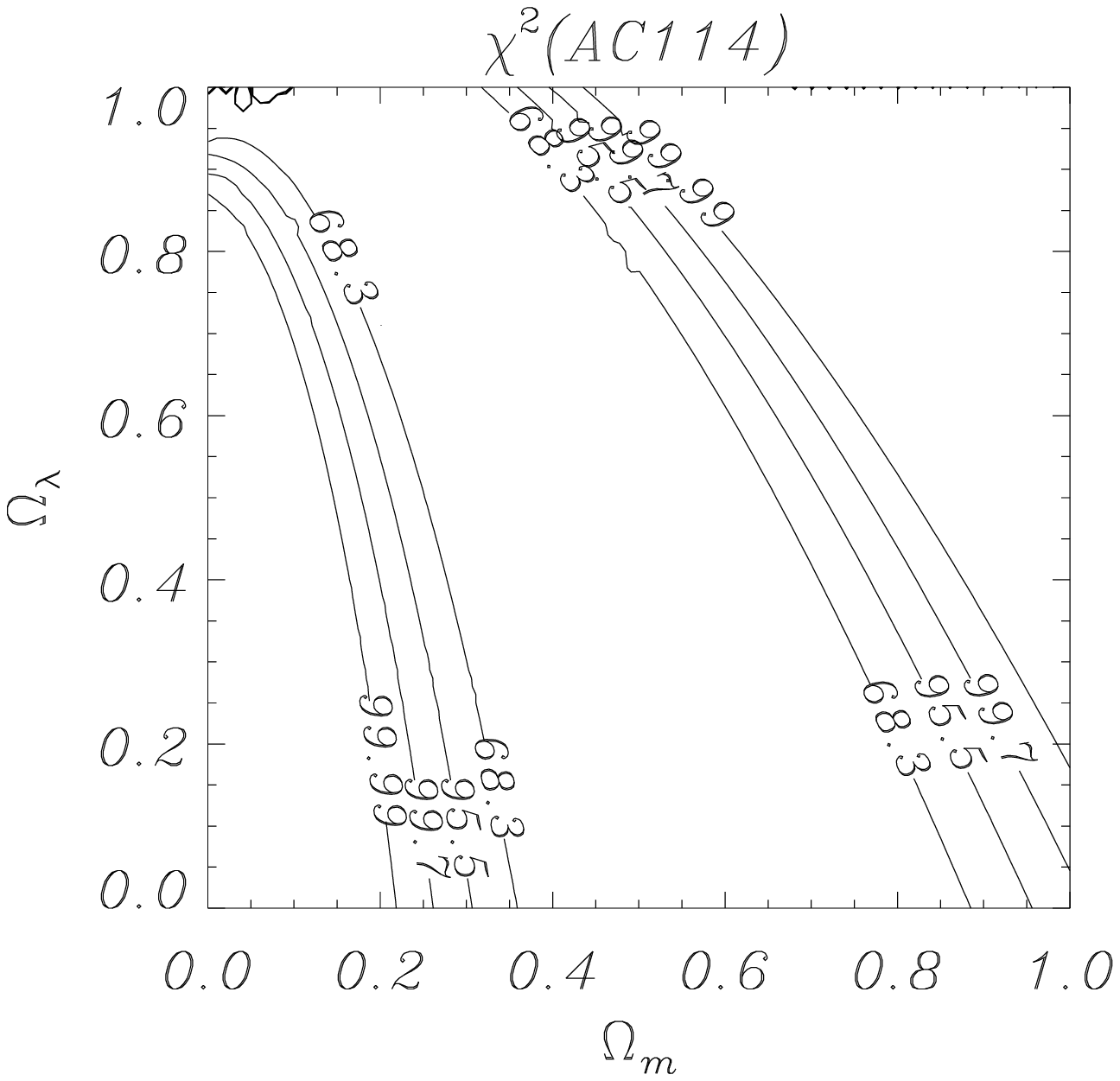}{4.4cm}{0.}{38}{38}{-170.}{0.}
\plotfiddle{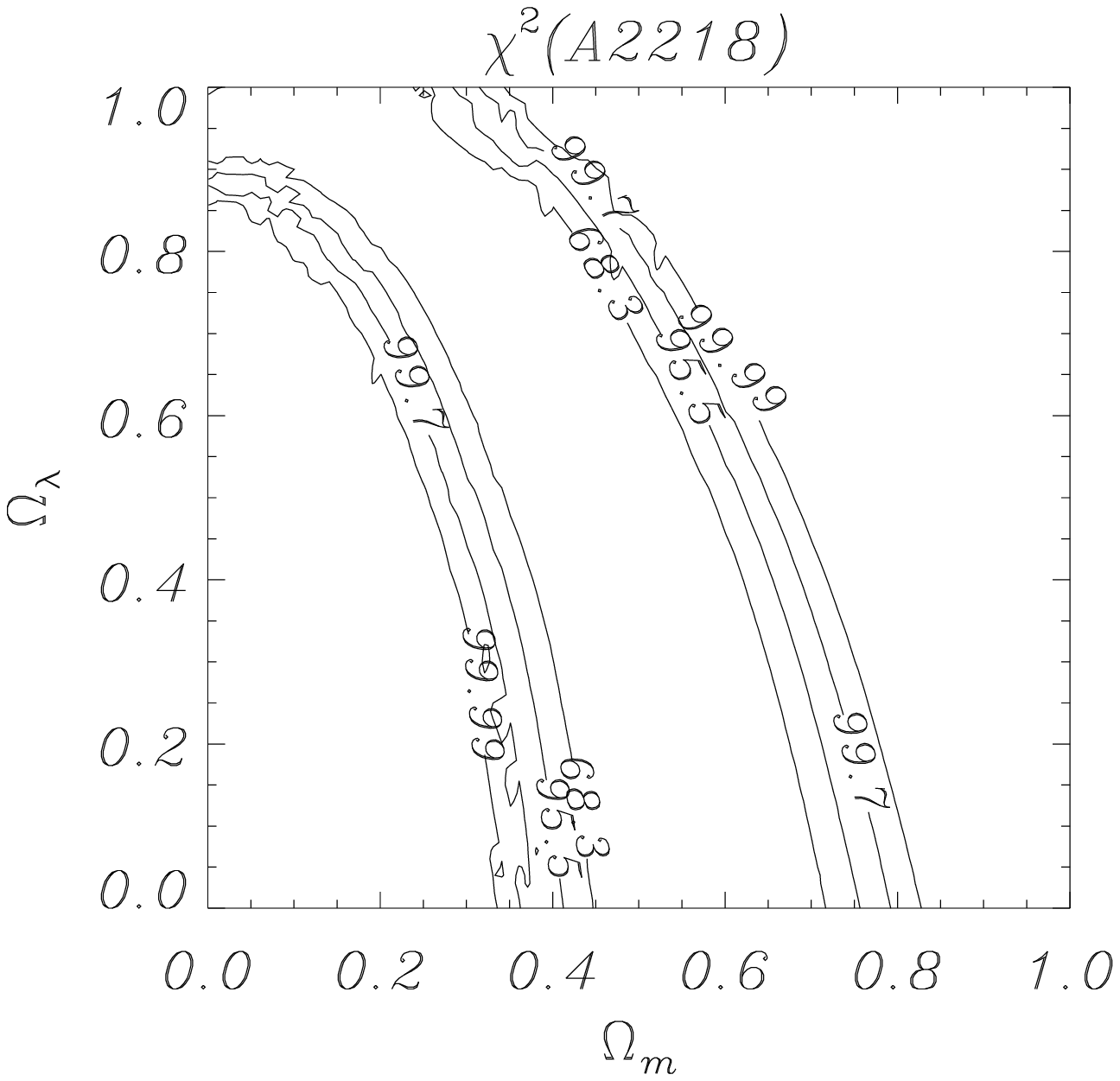}{0.cm}{0.}{38}{38}{10.}{24.}
\plotfiddle{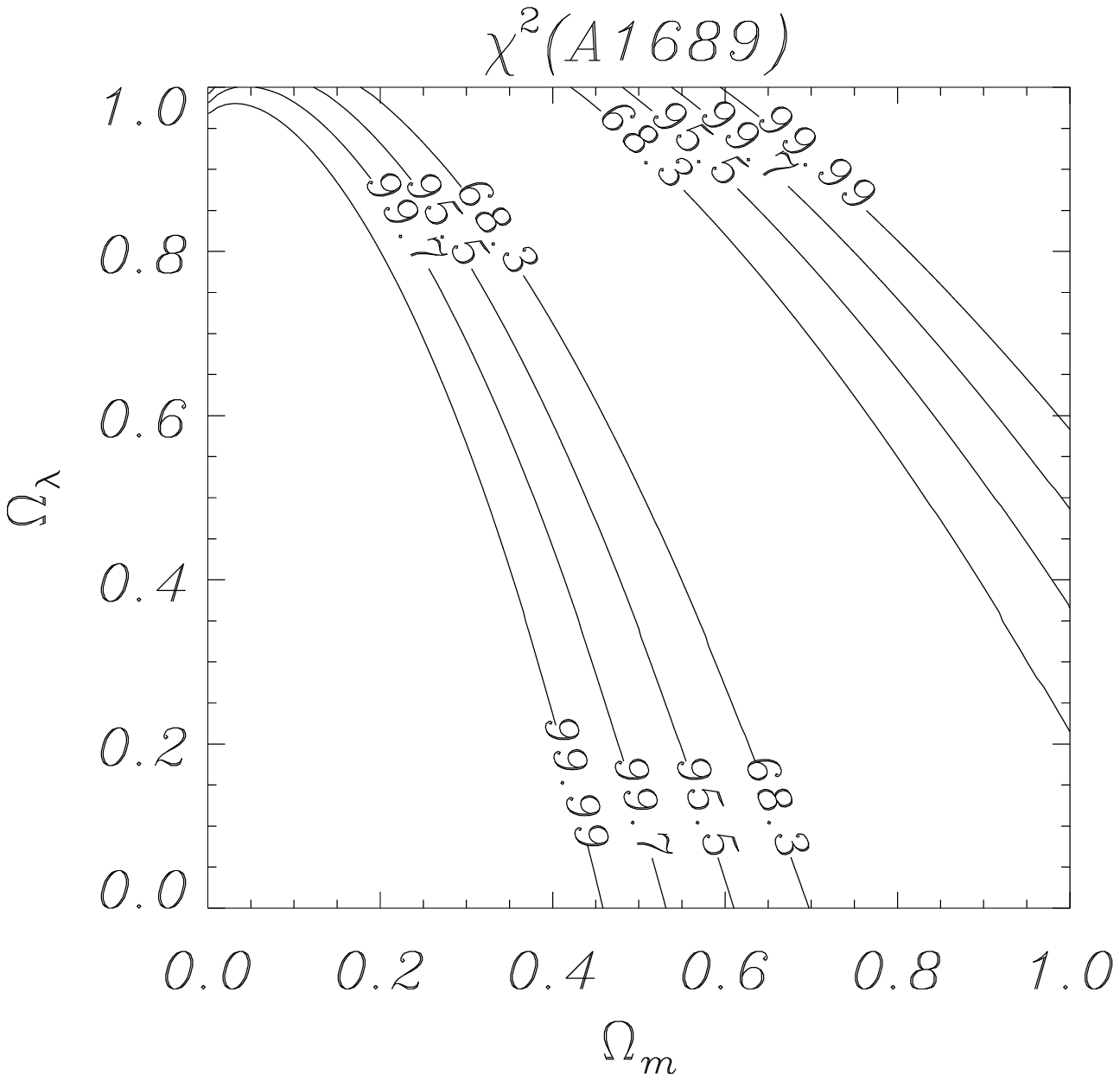}{4.1cm}{0.}{38}{38}{-170.}{0.}
\plotfiddle{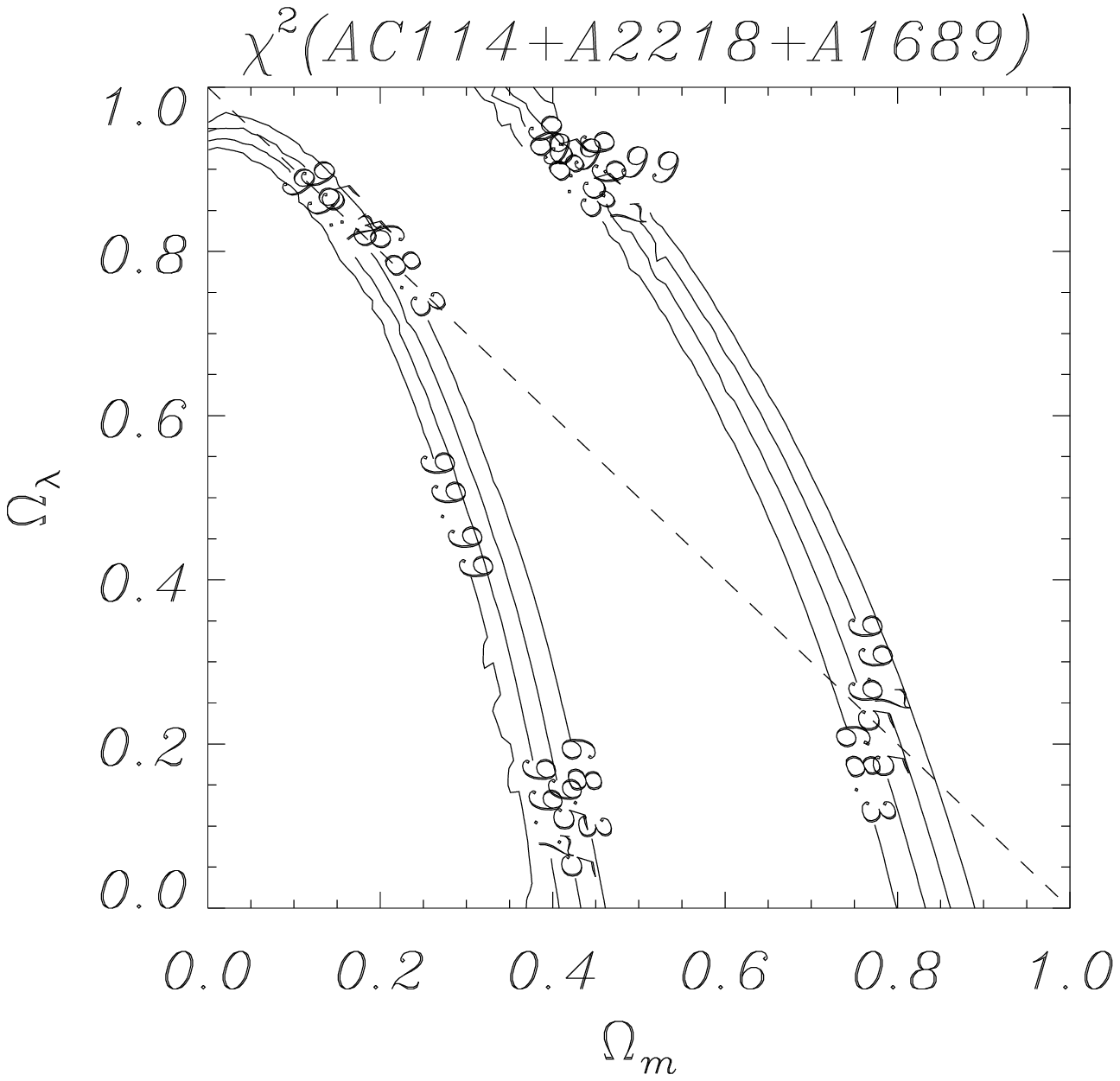}{0.cm}{0.}{38}{38}{10.}{24.}
\caption{$\chi^2$ confidence levels in the 
$(\Omega_{\mathrm{m}},\Omega_\lambda)$ plane obtained in the optimisation of 
the potential of the cluster-lenses AC114, A2218, A1689, and a
combined result of the three (dashed line: $\Omega_{\mathrm{m}}+
\Omega_\lambda=1$).}
\end{figure}

\section{Conclusion}

Combining the constraints from these 3 clusters lead to the Fig.~1 confidence 
levels. We obtain meaningful constraints compatible with a flat universe 
$(\Omega_{\mathrm{m}}+\Omega_\lambda=1)$. Since the exact degeneracy depends 
on the different redshift planes involved, results from other cluster lenses 
will further tighten the error bars. 
This test can be part
of a joint analysis to improve the precision on the parameters.

\end{document}